\documentclass[aps]{revtex4}
\usepackage{graphicx}

\def\bea{\begin{eqnarray}}
\def\eea{\end{eqnarray}}

\def\sqr#1#2{{\vcenter{\vbox{\hrule height.#2pt
      \hbox{\vrule width.#2pt height#1pt \kern#1pt
         \vrule width.#2pt}
      \hrule height.#2pt}}}}

\def\figloc#1#2 {
\begin{figure}\begin{center}
    \includegraphics[width=120mm]{fig#1.ps}
    \caption{ #2}
    \end{center}\end{figure}
}

\def\asinh {\rm asinh}

\begin{document}
\title{ Universal coordinates for Schwarzschild black holes  }

\author{W. G. Unruh}
\affiliation{ CIAR Cosmology and Gravity Program\\
Dept. of Physics\\
University of B. C.\\
Vancouver, Canada V6T 1Z1\\
~
email: unruh@physics.ubc.ca}

~

~

\begin{abstract}
A variety of historical coordinates in which the Schwarzschild metric is
regular over the whole of the extended spacetime are compared and the
hypersurfaces of constant coordinate are graphically presented. While the
Kruscal form (one of the later forms)  is probably the simplest, each of the others has some interesting
features. 
\end{abstract}

\maketitle

For years after  Schwarzschild\cite{schwarz} found a solution for spherically symmetric
metrics to Einstein equations,
\bea
ds^2= (1-{2M\over r})dt^2- {dr^2\over 1-{2M\over r}}
-r^2(d\theta^2+\sin(\theta)^2 d\phi^2)
\eea
 the status of the singularity at $r=2M$ (in
units where $c=1~G=1$) confused many, including Einstein\cite{eisenstaedt}. It was only in 1933, when
Lema\^itre\cite{lemaitre} found his coordinate transformation that he explicitly stated that that singularity in
the metric was an artifice introduced because of the coordinates that
Schwarzschild had used. It had already been recognized by Lanczos in 1922 that
the status of singularities in a  metric was unclear because singularities
could be introduced by making a singular choice of coordinates. However, the
application of this to the $r=2M$ singularity not appreciated.  
In 1921, both Gullstrand and Painleve\cite{PG} had found  new, spherically symmetric
solutions to Einstein's equation,
\bea
ds^2= (1-{2M\over r})d\tau^2-2\sqrt{2M\over r} dt dr
-r^2(d\theta^2+\sin(\theta)^2 d\phi^2)
\eea
In the following I will refer to this as the PG form of the metric.
  They, however, did not
recognize that this solution is  simply a  coordinate transformations of
 Schwarzschild's solution, nor did
they recognize the implication for the Schwarzschild singularity, believing
that coordinates themselves held physical significance. 

In the Kruskal\cite{SK}  paper, the claim is made that Kasner\cite{kasner} in 1921
showed that the $r=2M$ singularity was a just a coordinate singularity. This is not
true. Kasner embedded the Schwarzschild solution into a 6 dimensions (signature
4+2) flat spacetime but that embedding is singular at $r=2M$-- it covers only
the region $r>2M$.

In 1922,
Eddington\cite{eddington} found an explicit coordinate transformation which gave the metric
\bea
ds^2= (1-{2M\over r}) (d\tilde t+dr)^2 -2d\tilde t dr -2dr^2 -r^2(d\theta^2+\sin(\theta)^2
d\phi^2)
\eea
which is regular at $r=2M$, 
but did not recognize (or at least did not comment on ) the implication that
this had for the Schwarzschild singularity. (This coordinate transformation
and metric were rediscovered in 1954 by Finkelstein\cite{finkelstein} who certainly did recognize
that this implied that the Schwarzschild singularity was purely a coordinate
artifact. What is now called the Eddington-Finkelstein (EF)  form of the metric is
obtained from their form by replacing $t$ by $v=\tilde t+r$ to give
\bea
ds^2= (1-{2M\over r}) (dv)^2 -2dvdr  -r^2(d\theta^2+\sin(\theta)^2
d\phi^2) 
\eea
but this null form was never actually written down by either of them.)

In the following I will chose spatial units so that $2M=1$.
Thus the Schwarzschild metric becomes
\bea
ds^2= (1-{1\over r}) dt^2-{dr^2\over 1-{1\over
r}}-r^2(d\theta^2+\sin(\theta)^2 d\phi^2)
\eea

Both the PG metric and the EF metric are
coordinate transformations of each other, with a transformation is regular for
all values of $r>0$. In particular, if we take 
\bea
\tau=v -r+\sqrt{r} 
\eea
we turn the PG into the EF form of the  metric. 

In 1933, Lema\^tre, concerned about cosmological solutions to Einstein's
equations, introduced his form  of the Schwartzschild metric. He was
interested in the solution in which one embeds a Schwarzschild solution in a
De-Sitter universe, but also took the limit as the cosmological constant was
zero.
\bea
ds^2= d\tau^2 -{2M\over r(\sigma-t)} d\sigma^2 -r(\sigma-t)^2
(d\theta^2-\sin(\theta)^2d\phi^2)
\eea
where $r(\sigma-\tau)= {1\over 2M} \left({3\over 2}(
\sigma-\tau)\right)^{2\over 3}$ and $\tau$ is the same time coordinate as in the PG
form of the metric. Lema\^itre was the one  that showed, in passing that this
was simply a coordinate transformation of the PG metric, and that the PG
metric itself was just a coordinate transformation of Schwarzschild's form.

What is interesting about all three forms of the metric (PG,EF, and
Lema\^itre) is
that while they do demonstrate that the Schwarzschild singularity is a
coordinate artifact, and in all three, the metric is regular (has a well
defined inverse everywhere including at $r=2M$) they come in two forms.
We can define two possibilities for the coordinate transformation.
For the EF metric
\bea
t= u_\pm \pm (r+2M\ln\left({r-2M\over 2M}\right)\\
ds^2= (1-{2M\over r})du_\pm^2 \pm 2 du_\pm dr -r^2 (d\theta^2+\sin(\theta)^2
d\phi^2) 
\eea
For the PG metric
\bea
t = \tau\pm(\sqrt{2Mr} +2M\ln\left({ \sqrt{2Mr}-2M\over
\sqrt{2Mr}+2M}\right)
\\
ds^2= (1-{2M\over r})d\tau_\pm \pm\sqrt{2M\over r} d\tau_\pm dr -dr^2
-r^2(d\theta^2 +\sin(\theta)^2 d\phi^2)
\eea
and for the Lem\^itre metric, the PG transformation plus the extra
transformation
\bea
r\sqrt{r}=\tau_{\pm}  \pm \sigma_\pm\\
ds^2= d\tau_\pm - {1\over r}d\sigma_\pm -r^2(d\theta^2+\sin(\theta)^2 d\phi^2)
\eea
 
In all three cases the two solutions, labelled by $\pm$ are not the same
solution. While they are just coordinate transformations of each other for
$r>2M$, the spacetime covered is different for $r<0$. This can be most easily
seen by looking at the radial null geodesics. 

In the EF case, the null geodesics are
\bea
u_\pm=u_{0\pm}\\
u_\pm =u+{0\pm} \pm 2(r+2M\ln({r-2M\over 2M})
\eea
The first equation has a regular solution for $u_\pm$ for all values of $r$
while 
the second equation has $u_\pm$ go to $-\pm\infty$ as  $r\rightarrow 2M$. But
for the $u_+,r$ the first represent null rays which are travelling outward,
While the second is null rays which travel inward. Thus for the $u+$ the
ingoing null rays have no representation for $r<2M$. For $u_-$ it is the
opposite. The first represents null rays which travel inward, while the second
singular solution is null rays which travel outward. Thus for the $u_+,r$
coordinates, the region $r<0$ is where outward travelling null rays come from,
while for $u_-,r$ it is where ingoing null rays go to. Thus the regions $r<0$
are entirely different spacetimes in the two coordinate. 

Exactly the same occurs for the other two possibilities.
For PG coordinates, the null solutions are 
\bea
\tau_\pm = \tau_{0\pm} - (\pm\sqrt{2M\over r} +1)\\
\tau_\pm = \tau_{0\pm} - (\pm\sqrt{2M\over r} -1)
\eea
 while the second, irregular
solution is 
 if one changes the
sign of $v$ or $\tau$ one obtains a different solution of the Einstein
equations.
While outside $r>2M$ this new metric is simply a coordinate transformation of
the Schwarzschild, inside $r<2M$ it is not, the two forms cover different
spacetimes. 

The ingoing null geodesics in the EF metric are 
 give by $v$ constant, which is clearly regular for all values of $r>0$.
However the outgoing null rays obey
\bea
{dr\over dv}= {1\over2} (1-{2M\over r})\\
v-v_0 = 2\left(r+2M\ln\left({r-2M\over 2M}\right)\right)
\eea
with $v$ going to $-\infty$ as $r$ approaches $2M$. In the $u,r$ coordinates
obtained from this form by setting $u=-v$ (or making the coordinate
transformation from Schwartzschild of $t=u+r+2M\ln({r\over 2M}-1)$), the
outgoing null geodesics are $u$ constant, everywhere down to $r=0$ while the
ingoing null geodesics $u-u_0= 2(r+2M\ln({r\over 2M}-1))$ are singular as
$r\rightarrow 2M$. 

Similarly in the PG form of the metric, the outgoing null geodesics are given
by
\bea
\left( {dr\over d\tau}\right)^2 + 2\sqrt{2M\over r}{dr\over d\tau} -
(1-{2M\over r})=0
\eea
or
\bea
{dr\over d\tau}= -\sqrt{2M\over r} \pm 1
\eea
This has well behaved solutions at $r=2M$ for the minus  sign, but divergent
solutions there for the plus sign.  Again null geodesics going into the
horizon are well behaved through the horizon, while those coming out are badly
behaved. This is reversed with the other PG solution obtained when
$\tau\rightarrow -\tau$. 

Finally, the Lema\^itre form is more mysterious. Not only is the metric
diagonal but the metric looks completely regular at $r=2M$ (or rather
$\sigma-\tau = 3M$)
The null geodesics are given by
${d\sigma\over d\tau} =\pm {r(\sigma-\tau)\over 2M}$ However, writing this in
terms of the variable $r$ rather than $\sigma$ we obtain exactly the PG null
geodesics which we know are singular at $r=2M$.

Is there a set of coordinates for which the only singularities occur at r=0,
and in which the null geodesics are all regular at $r=2M$?  The answer is of
course yes, and the 
best known answer is the Kruskal-Szekeres form. However, such a coordinate
system was first   given by
Synge\cite{synge}
in 1950.

In the following I will choose units for my coordinates so that $2M=1$ so
factors of $2M$ do not have to be dragged along through all of the equations.

Write the Schwarzschild metric in terms of the proper distance  to the horizon
\bea
R= \int_1^r {dr\over \sqrt{1-{1\over r} }}= \sqrt{r-1} \sqrt r +
 \asinh(r-2M) =\sqrt{r-1} \left(\sqrt{r} + {\asinh(\sqrt{r-1})\over \sqrt{r-1}}\right) 
\eea
We have 
\bea
1-{1\over r} = 2{R^2\over r\left(\sqrt{r} +
{\asinh(\sqrt{{r}-1})\over
\sqrt{{r}-1} }\right)^2}
\eea

\bea
ds^2= F(r(R)) R^2 {1\over 4} dt^2 -dR^2 -r(R)^2 (d\theta^2+\sin(\theta)^2 d\phi^2)
\eea
where
\bea
F(r) ={ 4\over  r\left(\sqrt{r} + {\asinh(\sqrt{{r}-1})\over
\sqrt{{r}-1}}\right)^2}
\eea
The function $F(r(R))$ looks singular at $r=1$ but is not.  $\sqrt{r}$ is analytic for
$r>0$.  
The function 
${\asinh(\sqrt{{r}-1})\over \sqrt{{r}-1}}$ is also an analytic
function of $r$ everywhere for $r>0$. It is an even funtion in the argument $\sqrt{{r}-1}$
and is thus analytic in $r$ for $r>0$.$F(r)$ is also monotonic in $r$ and thus$R^2$ is an
analytic monotonic function of $r$ for $r>0$ and thus so is $r(R)$.

Also $F(r=1)=1$ and we can thus write the metric as
\bea
ds^2= (F(r(R))-1) R^2 {1\over 4}dt^2 + R^2 dt^2-dR^2-r(R)^2 (d\theta^2+\sin(\theta)^2
d\phi^2)
\eea
Now defining
\bea 
T=R\sinh(t/2)
\\
\xi=R \cosh(t/2)
\eea
and thus $R^2= \xi^2-T^2$, 
we have the regular metric
\bea
ds^2= (F(r(R))-1) (\xi dT-Td\xi)^2 + dT^2-d\xi^2 -r(\sqrt{T^2-\xi^2})^2
(d\theta^2+\sin(\theta)^2 d\phi^2)
\eea
This metric is singular for $T^2-\xi^2=\pi$ (which corresponds to $r=0$) but is regular everywhere
else. This is the Synge form of the Schwarzschild metric, the first of the
metric forms whose coordinates cover all of the analytically extended  spacetime (all geodesics
either end in a genuine singularity, corresponding to one of  the $r=0$
singularities, or extend to
infinity.)
Note also that the lines of $\xi,\theta,\phi$ constant are not necessarily
timelike lines. for $\xi$ sufficiently large and $r$ sufficiently small,
$F((r)-1)\xi^2 +1$ can be negative of $r<1$ and thus the line becomes
spacelike.

The Szekeres-Kruskal metric can be formed in the same way. Define
\bea
ds^2= G(r(\rho))( \rho^2\alpha^2dt^2-d\rho^2) -r(\rho)^2(d\theta^2+\sin(\theta)^2 d\phi^2)
 \eea
where $\alpha$ is a constant.
This leads to 
\bea
{d\rho\over dr}=\alpha {\rho\over 1-{1\over r}}
\eea
or
\bea
\rho= ({r}-1)^{\alpha} e^ \alpha r
\eea
Choosing $\alpha={1\over 2}$ we have
\bea
ds^2=  e^{-{r(\rho)\over 2}} {1\over r} (\rho^2 ({dt\over
4M})^2-d\rho^2) - r(\rho)^2 (d\theta^2+\sin(\theta)^2 d\phi^2)
\eea
Defining
\bea
\tau=\rho \sinh({t\over 2})\\
\chi=\rho \cosh({t\over 2})
\eea
we get the Szekeres/Kruskal metric
\bea
ds^2=  e^{-r\over 2} {1\over r}(d\tau^2-d\chi^2)
-r(\tau^2-\chi^2)^2(d\theta^2+\sin(\theta)^2 d\phi^2)
\eea

There is another way of arriving at the same result. Writing the EF metric 
\bea
ds^2= (1-{1\over r}) du_{\pm}^2 \pm 2du_{\pm} dr
--r^2(d\theta^2+\sin(\theta)^2 d\phi^2)
\eea
with
\bea
u_{\pm}= t\pm (r+\ln({r}-1)
\eea
to give
\bea
{r}-1= e^{u_+-u_- -2r \over 2} {1\over r} 
-r^2(d\theta^2+\sin(\theta)^2 d\phi^2)
\eea
to give
\bea
ds^2= e^{-{r}} {1\over r} (e^{u_+\over 2} du_+)( e^{-{u_-\over 2} }
du_-) -r^2(d\theta^2+\sin(\theta)^2 d\phi^2)
\eea

Defining $U_\pm =\pm 4M e^{\pm u_{\pm}}$ and
\bea
\tau=(U_++U_-)/2\\
\chi=(U_+-U_-)/2
\eea
we obtain exactly the Szekeres-Kruskal metric obtained before.

This second procedure for finding the SK coordinates also allows us  to carry out the same procedure for the PG metric
Defining
\bea
\tau_\pm= t\pm(  2\sqrt{r}
+\ln\left( {\sqrt{r}-1\over \sqrt{r}+1}\right) 
 \eea
we have
\bea
d\tau_\pm^2 \pm {\sqrt{1\over r}\over 1-{1\over r}}d\tau_\pm dr
-r^2(d\theta^2+\sin(\theta)^2 d\phi^2)
\eea
In terms of these "times" we have
\bea
ds^2= -{{r}-1\over 4{r}}(d\tau_+^2+d\tau^2_-) + {({r
})^2-1\over {r}} d\tau_+ d\tau_- -
r^2(d\theta^2+\sin(\theta)^2 d\phi^2)
\eea
Defining $\Xi_\pm= e^{\tau_\pm/2}$ and $y=\sqrt{r}$, we have
\bea
ds^2=(2M^2) \left[ -{e^{-4y} (y+1)^4\over y^2}(\Xi_+^2d\Xi_-^2+\Xi_-^2
d\Xi_+^2) + -2e^{-2y}{(y^2+1)(y+1)^2\over y^2}d\Xi_+ d\Xi_- \right]
\eea
where $r(\Xi_+\Xi_-)$ is defined by
\bea
\Xi_+\Xi_- = {y-1\over y+1}e^{2y} 
\eea

This is again a regular metric everywhere where $r>0$ ($\Xi_+\Xi_- > -1)$.
 It retains the feature of the PG metric that the surfaces $\Xi_+=$const
or $\Xi_-=$const are flat spacelike surfaces-- ie it foliates the
extended Schwarzschild spacetime with a series of intersecting  flat spatial slices. 

Another  interesting metric is obtained by taking the Lema\^itre metric,
obtained from the Schwarzschild by the coordinate transformation
\bea
\tau= t+\int {sqrt{1/r}\over 1-{1\over r}} dr= \sqrt{r} +\ln\left({\sqrt{r}-1\over
\sqrt{r}+1}\right)\\
\sigma= \tau +{2\over 3} {r\sqrt{r}} 
\eea
which gives the metric
\bea
ds^2= d\tau^2 - {d\sigma^2} \over r -r^2(d\theta^2+\sin(\theta)^2 d\phi^2)
\eea
where $r= \left({3\over 2}(\sigma-\tau)\right)^{3\over 2}$.

Again, taking $\tau\rightarrow -\tau$ gives another solution which covers a
different sector of the spacetime than does the above metric. Taking
$\tau_\pm$ as two coordinates leads to the same metric as the above extended
PG metric. However we can also take
\bea
\Sigma_\pm= exp( \left({1\over 2}\left({2\over 3} r^{3\over 2} +\sqrt{r}
+\ln\left({\sqrt{r}-1\over\sqrt{r}+1}\right) \pm t\right)\right)
\eea
from which we find
\bea
\Sigma_+\Sigma_-= {\sqrt{r}-1\over \sqrt{r}+1} e^{{2\over 3} r^{3\over 2}
+\sqrt{r}}
\\
{\Sigma_+\over \Sigma_-}= e^t
\eea
and the metric becomes
\bea
ds^2= e^{-{{2\over 3} r^{3\over 2}
+\sqrt{r}}}(\sqrt{r}+1)^2 (\Sigma_+^2 d\Sigma_-^2 +\Sigma_-^2d\Sigma_+^2 
-2d\Sigma_-d\Sigma_+) -r^2 (d\theta^2 +\sin(\theta)^2 d\phi^2) 
\eea
In this case the surfaces of either $\Sigma_+$ or $\Sigma_-$ constant are
timelike surfaces and  the lines in those surfaces of $\theta$ and
$\phi$ constant are time-like geodesics in the Schwarzschild metric.

As a final example, we can look at a coordinate system related to the global
embedding of the Schwarzschild metric found by Fronsdale. 

Define the funtion $\hat R$ by
\bea
\hat R^2= 4(1-{1\over r})
\eea
$\hat R$ runs from $-\infty$ ($r=0$)  to 0 ($r=\infty$).
Then we can write
\bea
ds^2&=& \hat R^2 \left({dt\over 2}\right)^2 - d\hat R^2 -{1+r+r^2+r^3\over r^3}
dr^2 - r^2(d\theta^2+\sin(\theta)^2 d\phi^2)\\
&=& \hat R^2 \left({dt\over 2}\right)^2 - d\hat R^2  -\left({1+\left({1\over(1-({\hat R^2\over
4}}\right)^4}\right)\hat R^2
d\hat R^2- {1\over 1-{\hat R^2\over 4}}(d\theta^2+\sin(\theta)^2 d\phi)^2
\eea

As before, define 
\bea
\Theta=\hat R \sinh({t\over 2})\\
Y= \hat R \cosh({t\over 2})
\\
\hat R^2 =Y^2-\Theta^2
\eea

This gives
\bea
ds^2= d\Theta^2-dY^2 -(YdY-\Theta d\Theta)^2 \left( 1\over
1-{{Y^2-\Theta}^2\over 4}\right)^4 - {1\over 1-{Y^2-\Theta^2\over 4}}
(d\theta^2+\sin(\theta)^2 d\phi^2)
\eea

These are related to the global embedding of the Schwarzschild
metric in a 6-dimensional flat spacetime, first suggested by
Fronsdale\cite{fronsdale}. Defining the Z coordinate by
\bea
Z= \int^r {{r'}^2+r'+1\over {r'}^3} dr'
\eea
the metric becomes
\bea
ds^2= d\Theta^2-dY^2-dZ^2 -dr^2-r^2(d\theta^2+\sin(\theta)^2 d\phi^2) 
\eea
with the above definition of $\Theta,Y,Z$ as functions of $t,r$ giving the
embedding functions of the 4 dimensional surface in the 6 dimensional flat
spacetime. 

\section{Relations between coordinates}
Since the SK coordinates are the most standard, let us compare the 
other two coordinate systems to the PK coordinates graphically. 

Let us first look at the generalised PG coordinates to the SK coordinates. 
The extended PG coordinate  surfaces of constant $Xi_\pm$  to those of the SK
coordinates. Using the SK coordinates $U=\tau-\rho$ and $V=\tau+\rho$ we have
\bea
{\Xi_+\over\Xi_-}= e^{t\over2M} = {V\over U}\\
\Xi_+\Xi_- = {\sqrt{r\over 2M} -1\over\sqrt{r\over 2M} +1} e^{2\sqrt{r\over 2M}}\\
UV= ({r\over 2M}-1)  e^{r/2M}
\eea
Ie, $\Xi_+\Xi_-$is a function of $UV$ given parametrically by the last two
equations.

The diagram indicates the graph of constant $\Xi_+$ and $\Xi_-$  spacelike
hyperspace's for a few values of each. 

Note that as $r\rightarrow \infty$, both $\Xi_+$ and $\Xi_-$ (for suitable
values) asymptote to the same line. in the $UV$ plane. Ie, the $\Xi_+,\Xi_-$
coordinates become degenerate as $r\rightarrow \infty$.1

\figloc{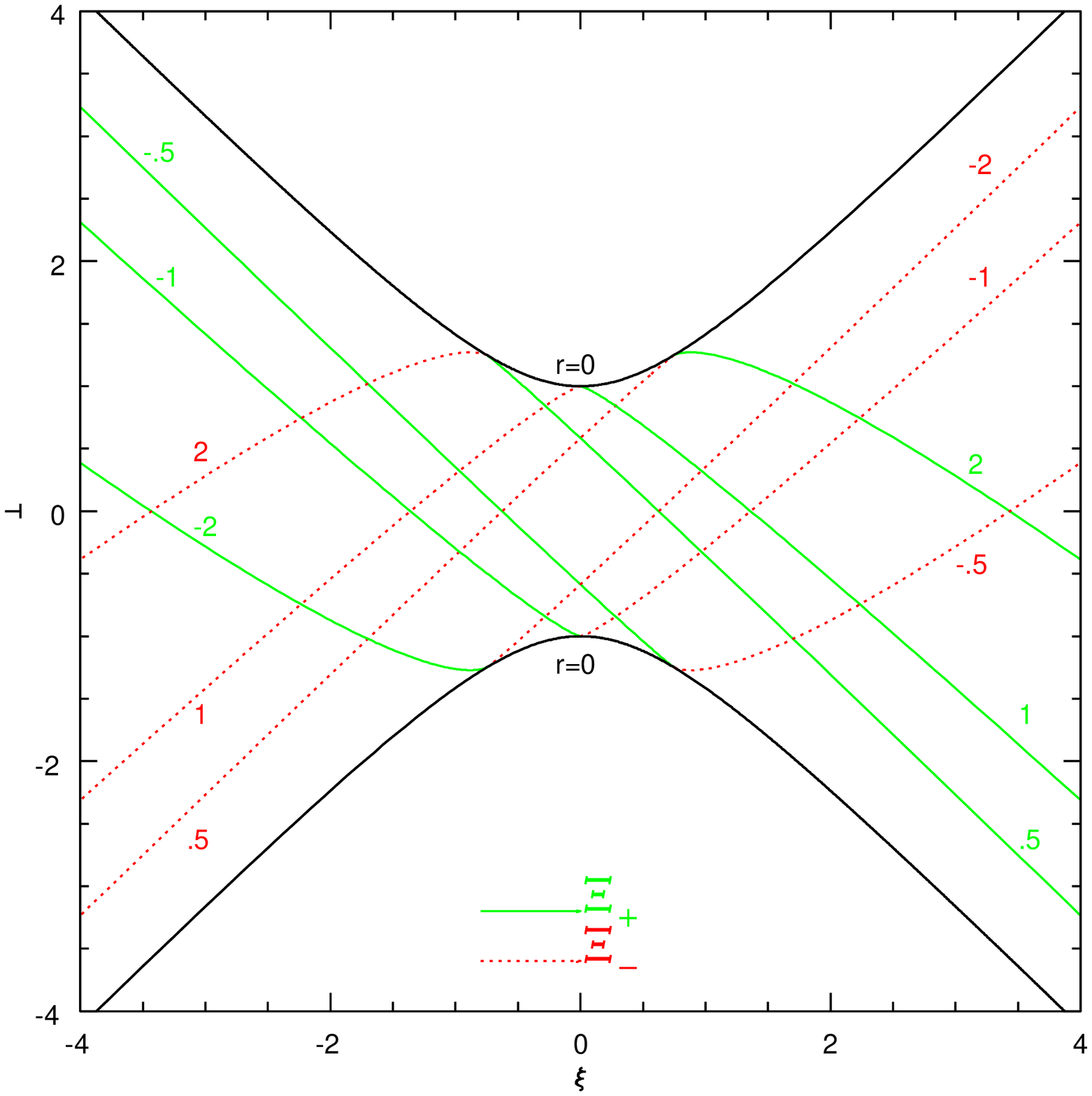}{The $\Xi$ constant coordinate surfaces in the Kruskal coordinates.
Each of those surfaces is a flat spatial slice. All begin at the r=0
singularity and go out to infinity. Note that both the $\Xi_+$ and the $\Xi_-$
constant surfaces are spatial surfaces. }

Then the Synge coordinates are plotted  vs the SK coordinates. The surfaces of
constant Synge time $T$  are given in terms of the SK coordinates
parametrically  by

\bea
{V+U\over 2}(T)&=& {T e^{r\over 2}\over\sqrt{r} + {\asinh(\sqrt{r-1})\over
\sqrt{r-1}}}\\
{V-U\over 2}(T) &=& \sqrt{ ({V+U\over 2})^2 +(r-1) e^r}
\eea
where $r$ must be large enough that $V-U\over 2$ is real. 

The $\xi$ coordinate constant surfaces are given by 
\bea
{V-U\over 2}(\xi)= {\xi e^{r\over 2}\over\sqrt{r} +
{\asinh(\sqrt{r-1})\over\sqrt{r-1}}}\\
{V+U\over 2}(\xi)= \pm\sqrt{ ({(V-U)\over 2})^2 -(r-1)e^r}
\eea
where the parameter $r$ is chosen small enough so that ${V+U\over 2}(\xi)$ is
real. 

In figure 2 we have the plot of the $T$ and $\xi$ constant surfaces in the SK
coordinates. 

\figloc{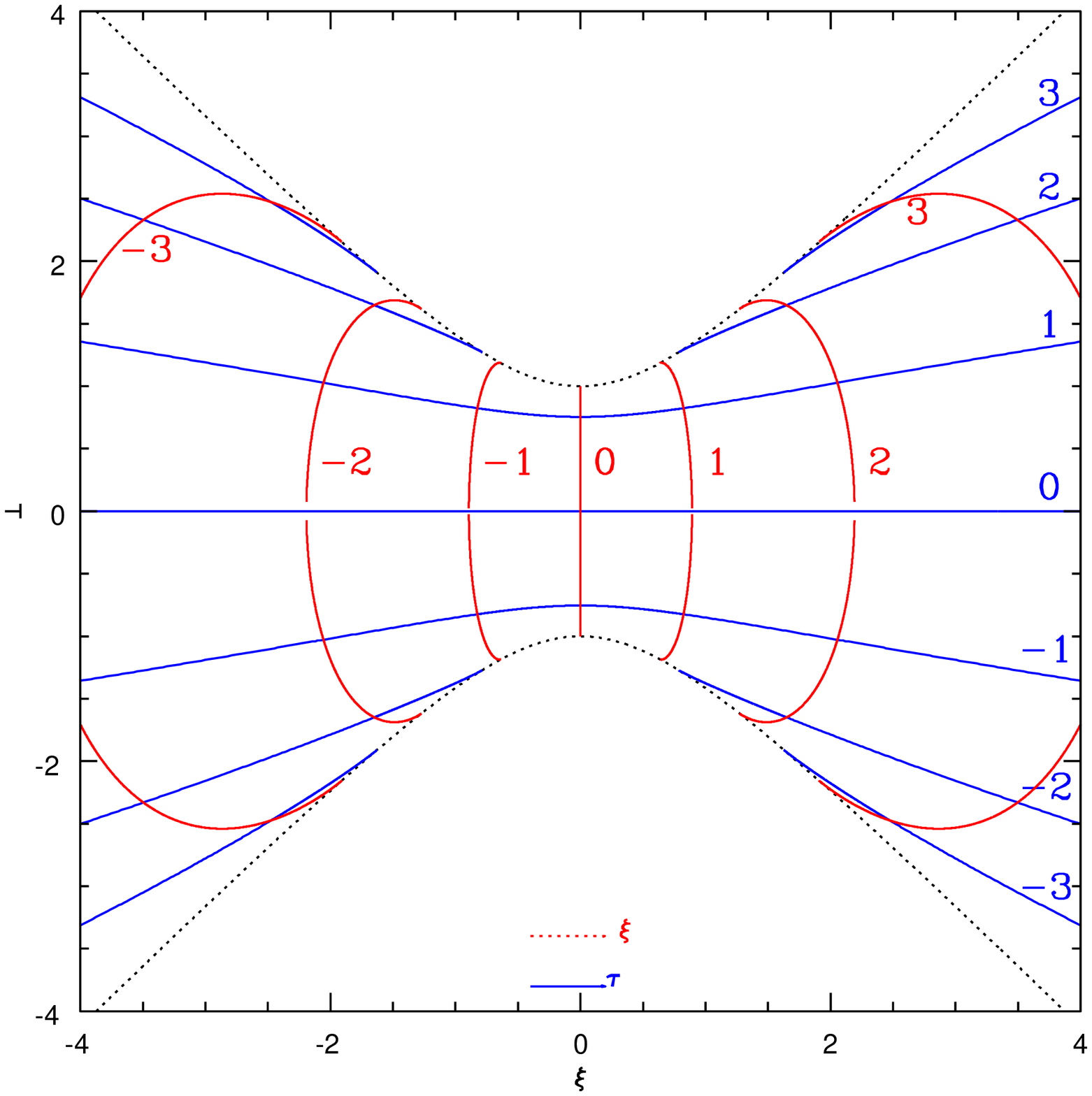}{ The $T,\xi$ constant coordinate surfaces plotted in Kruskal
coordinates. Note that while the $T$ constant hypersurfaces are spacelike
hypersurfaces, the $\xi$ constant one as not everywhere timelike. In
particular near and within the horizon these surface become timeline for large
enough values of $\xi$.}

The Lema\^itre coordinates are interesting because they look, at first, as
though they are regular coordinates already which cover the whole spacetime.
The metric
\bea
ds^2= d\tau^2 -{1\over r(\sigma-t)} d\sigma^2 -r(\sigma-t)^2
(d\theta^2-\sin(\theta)^2d\phi^2)
\eea
looks regular everywhere.except at $r=0$ of $t=\sigma$.
But if we look at the null geodesics
\bea
{d\sigma\over d\tau}=\pm \sqrt(r(\tau-\sigma))=\pm ({3\over 2}(\sigma-\tau))^{1\over 3}
\eea
we find for the + sign, taking $z=\sigma-\tau$ that
\bea
{dz\over d\tau}= \pm({3\over 2}z)^{1\over 3} -1
\eea
The RHS goes to 0 when $z={2\over 3}$ and $\tau$ goes to $\infty$ if we take
the $+$ sign in the equation for $z$. 	Ie, the null geodesics coming out of
the black hole come from $\tau\rightarrow-\infty$. Had one taken the other
solution ( with $\tau\rightarrow-\tau$) for the Lema\^itre metric, it would be
the ingoing null geodesics which would have terminated at $r=1$. Ie, again
the Lema\^itre coordinates cover only a part of the complete spacetime. The
extended Lema\^itre coordinates ($\Sigma_\pm$) do cover the whole of the
spacetime. 
 
\figloc{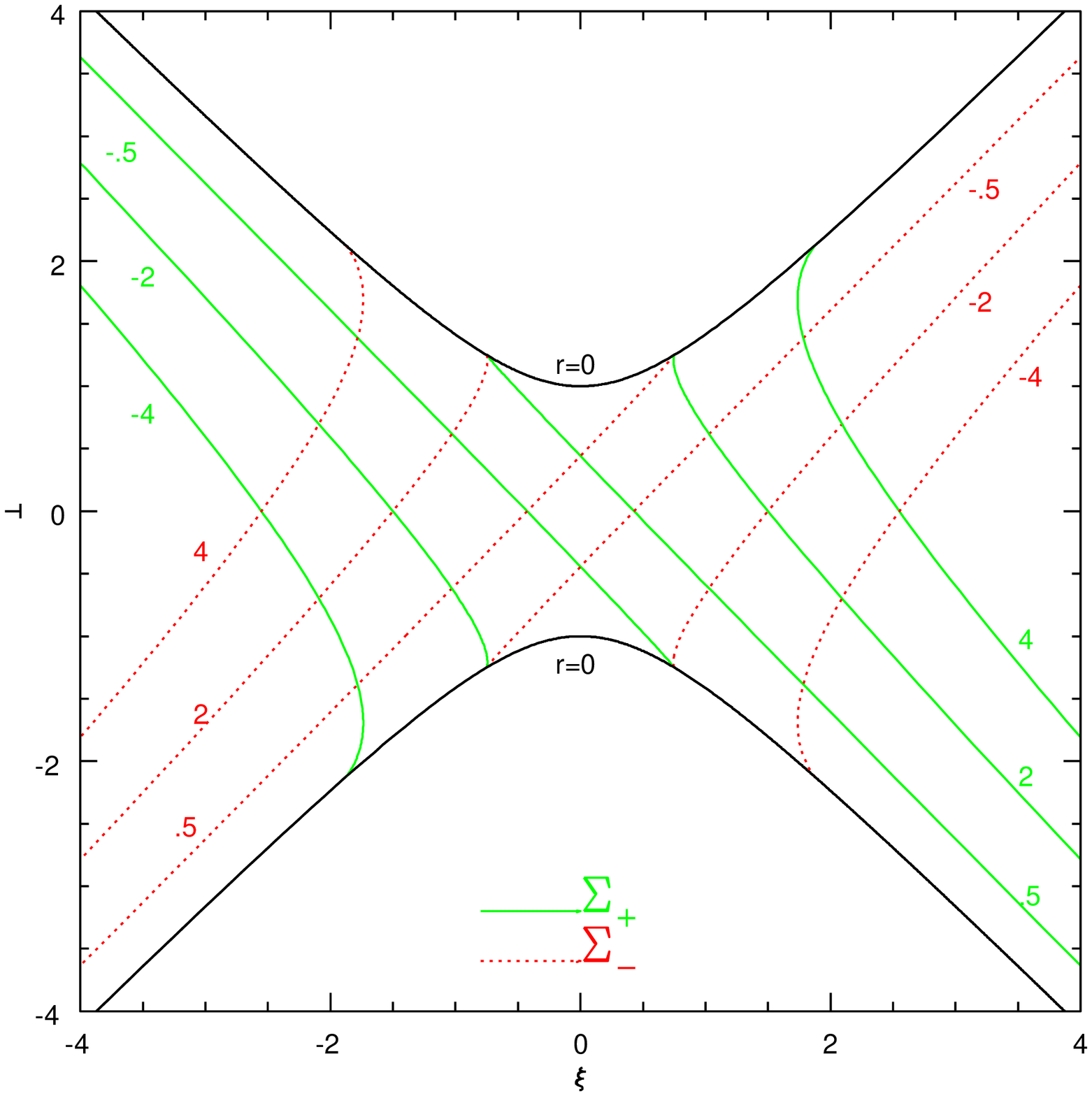}{The Lemaitre extended coordinates plotted on the SK extended
coordinates. }

From the two graphs, of the extended PG coordinates, and the extended
Lema\^itre coordinates, we can see the problem with the original Lema\^aitre
coordinates. The latter are essentially using the $\Xi_-$ and the $\Sigma_-$
coordinates. the problem with these is they become degenerate along the
past horizon, where both are equal to zero. Ie, these ( and the original
Lema\^itre coordinates which are the logarithm of these coordinates)
coordinates do not cover the past horizon. However, if we choose for example
the $\Sigma_+$ and the $\Xi_-$ coordinates, these do cover the whole of the
extended spacetime, with no degeneracies. 
We have
\bea
\Sigma_+\Xi_- = {\sqrt{r}-1\over\sqrt{r}+1} e^{\sqrt{r}({1\over 3}r+1)}\\
{\Sigma_+\over \Xi_-}= e^t e^{{1\over 3}r\sqrt{r}}
\eea
or
\bea
{\sqrt{r}(r+2)\over 2(r-1)}dr= {d\Sigma_+\over \Sigma_+} +{d\Xi_-\over \Xi_-} \\
dt +{1\over 2} \sqrt{r} dr = {d\Sigma_+\over \Sigma_+}-{d\Xi_-\over \Xi_-}
\eea
to give 
\bea
ds^2= {\sqrt{r}+1\over (r+1)^2} \left[ (\sqrt{r}+1)e^{-(r/3+2)\sqrt{r}} (\Xi_-^2
d\Sigma_+^2- \Sigma_+^2 d\Xi_-^2) +4d\Xi_- d\Sigma_+\right] +r^2
(d\theta^2+\sin(\theta)^2d\phi^2
\eea
This shares with the original Lema\^itre coordinates that each of the $\Sigma$
constant hypersurfaces are flat three dimensional spatial metrics, while each
of the $\Xi,\theta\phi$ constant lines are timelike geodesics which have zero
velocity at infinity. Unlike the original Lema\^itre coordinates however, they
cover the whole of the analytic extension of Schwarzschild spacetime. They are
thus just as simply married to the flat Robertson Walker dust universe model
as were the original Lemaitre coordinates.

Finally, using the Fronsdale coordinates $\Theta,Y$ we plot the $\Theta$
constant and $Y$ constant 
hypersurfaces. Note that these $\Theta$ constant hypersurfaces surfaces do not
run into the $r=0$ singularity. On the other hand, all of the $Y$ constant
lines originate at $T=\pm 1,~\xi=0$ points on the singularity, with the $Y$
constant lines only being timelike for certain values of $Y<2$ and only for
certain values of $\Theta$. Ie, the $Y$ constant coordinate in these ``Fronsdale"
coordinates is very badly behaved near the $r=0$ singularity while the
$\Theta$ const. coordinate surfaces are nicely behaved.

\figloc{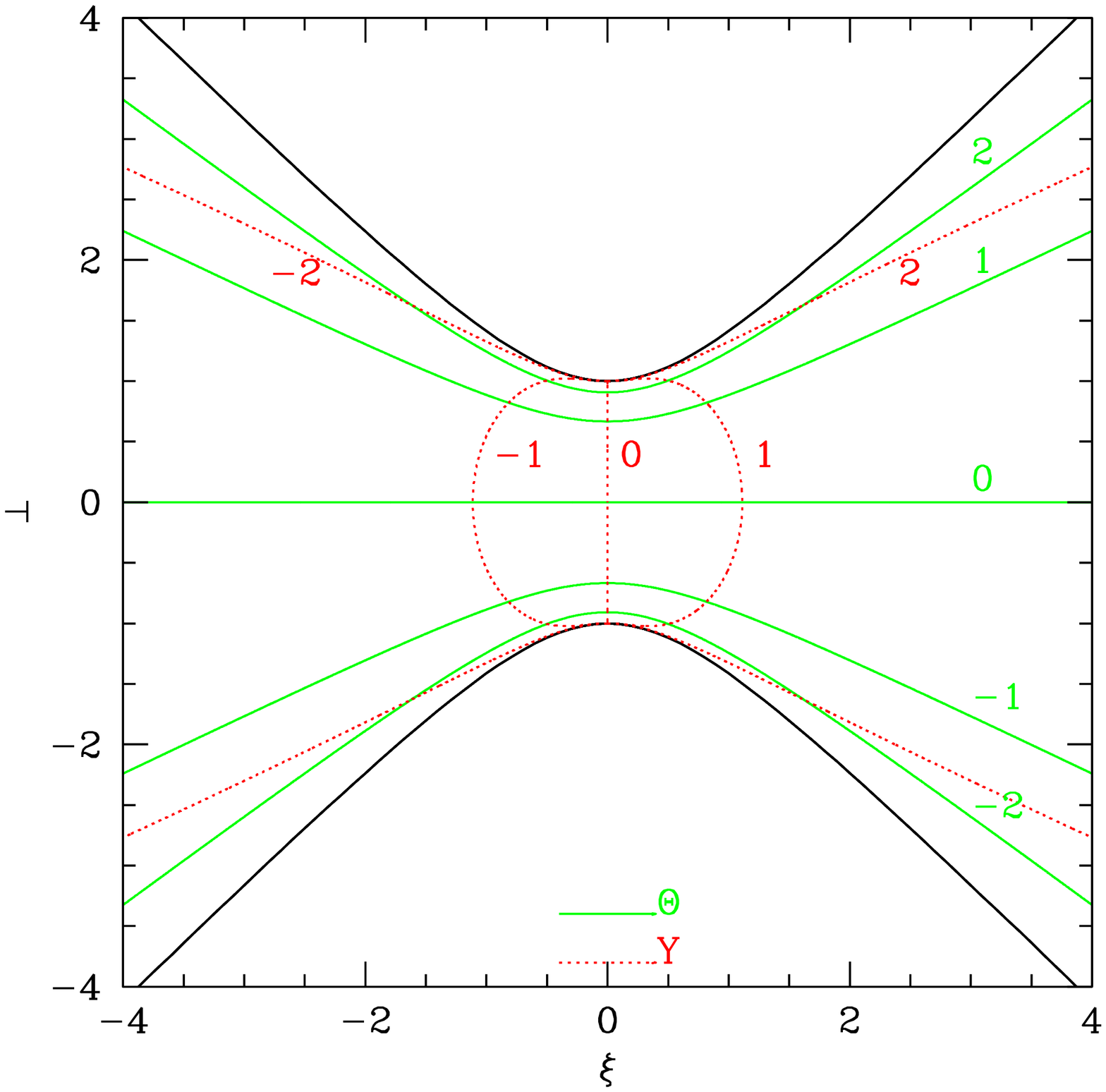}{The $\Theta$ and $Y$  constant hypersurfaces for the Fronsdale embedding of
Schwarzschild into a flat 6 dimensional spacetime, While the $Y$ constant
coordinates seem to hit the $r=0$ singularity are various points, those
surfaces actually skirt (as spacelike surfaces) extremely close to the
singularity before finally all hitting it at the same point.}

\end{document}